\documentclass[fleqn,usenatbib]{mnras}

\usepackage{newtxtext,newtxmath}

\usepackage[T1]{fontenc}

\DeclareRobustCommand{\VAN}[3]{#2}
\let\VANthebibliography\thebibliography
\def\thebibliography{\DeclareRobustCommand{\VAN}[3]{##3}\VANthebibliography}

\usepackage{graphicx}	
\usepackage{amsmath}	

\title[The Mass Budget Necessary to Explain `Oumuamua as a Nitrogen Iceberg]{The Mass Budget Necessary to Explain `Oumuamua as a Nitrogen Iceberg}

\author[A. Siraj \& A. Loeb]{
A. Siraj$^{1}$\thanks{E-mail: amir.siraj@cfa.harvard.edu}
and A. Loeb $^{1}$\thanks{E-mail: aloeb@cfa.harvard.edu}
\\
$^{1}$Department of Astronomy, Harvard University, 60 Garden Street, Cambridge, MA 02138, USA\\
}

\date{Accepted XXX. Received YYY; in original form ZZZ}

\pubyear{2021}

\begin{document}
\label{firstpage}
\pagerange{\pageref{firstpage}--\pageref{lastpage}}
\maketitle

\begin{abstract}
Recently, a nitrogen iceberg was proposed as a possible origin for the first interstellar object, 1I/2017 U1, also known as `Oumuamua. Here, we show that the mass budget in exo-Pluto planets necessary to explain the detection of `Oumuamua as a nitrogen iceberg chipped off from a planetary surface requires a mass of heavy elements exceeding the total quantity locked in stars with 95\% confidence, making the scenario untenable because only a small fraction of the mass in stars ends in exo-Plutos.
\end{abstract}




\section{Introduction}

1I/2017 U1, also known as `Oumuamua, was the first interstellar object discovered in the solar system \citep{2017Natur.552..378M, 2018Natur.559..223M}. Recently, \cite{2021JGRE..12606706J} and \cite{2021JGRE..12606807D} proposed that `Oumuamua may be an $\mathrm{N_2}$ iceberg chipped off from the surface of an exo-Pluto by an impact during a period of dynamical instability.

Here, we demonstrate that the mass density required to produce enough such fragments to make the detection of `Oumuamua a likely event is unreasonably large, thereby strongly disfavoring the nitrogen iceberg scenario. In Section \ref{sec:mb}, we explore the mass budget necessary to produce a population of $\mathrm{N_2}$ icebergs that would explain the detection of `Oumuamua, and in Section \ref{sec:d} we discuss our main results. 


\section{Mass budget}
\label{sec:mb}

In the $\mathrm{N_2}$ model, the initial mass of `Oumuamua is $m = 2.4 \times 10^{11} \mathrm{\; g}$ \citep{2021JGRE..12606706J, 2021JGRE..12606807D}, based on the inferred dimensions of $45 \mathrm{\; m} \times 44 \mathrm{\; m} \times 7.5 \mathrm{\; m}$ at the time of observation, combined with the optimistic assumption that it was ejected from its parent system only $\tau \approx 0.45 \mathrm{\; Gyr}$ ago. Given the $\mathrm{N_2}$ surface depth $D_{N2}$ of a few km, each Pluto contains $(m_{N2} / m_{p}) = 3 (D_{N2} \rho_{N2} / D_{p} \rho_p) \sim 0.5\%$ of its mass in nitrogen \citep{1998ASSL..227..655C, 2016Natur.534...82M, 2021JGRE..12606807D}. The parameters of the Pan-STARRS survey  \citep{2018ApJ...855L..10D} combined with the fact that no new `Oumuamua-like objects have been detected in the past few years, imply that the abundance of `Oumuamua-like objects is $n = 0.1^{+0.457}_{-0.097} \mathrm{\; AU^{-3}}$, quoted here with the $95\%$ Poisson error bars for a single detection. The local density of stellar mass is $\rho_{\star} = 0.04 \; \mathrm{M_{\odot} \; pc^{-3}}$ \citep{2017MNRAS.470.1360B}. Hydrogen and helium make up a negligible fraction of Pluto's mass but $(1 - f_Z) = 98.7\%$, of the mass of solar metallicity material \citep{2009ARA&A..47..481A}. We conservatively adopt the dynamical fudge factor of $\epsilon \sim 0.2$, which is twice the factor implied in \cite{2021JGRE..12606807D}, since such a multiplicative factor exceeds the maximum bounds of adjusting the underlying factors; namely, twice the product of $0.6$ for the upper bound on the fraction of excavated material that is pure $\mathrm{N_2}$ (between Plutos and Gonggongs), since it was determined for a hypothetically thicker ice layer, $0.18$ for surviving collisions and sublimation, and $0.8$ for ejection. Adopting perfect efficiency, $\epsilon = 1$, would still present a significant challenge for the $\mathrm{N_2}$ hypothesis. We denote by $\eta$ the fraction of the total mass in nitrogen icebergs that lies within a logarithmic bin around the inferred mass of `Oumuamua. We choose the upper limit for $\eta$, representing the most conservative value possible, $\eta \sim 1$, corresponding to all of the nitrogen icebergs being the size of `Oumuamua. We note that a more realistic assumption of a scale-free fragment size distribution with an equal mass per logarithmic size bin \citep{2020arXiv201114900S} spanning ten orders of magnitude would yield $\eta \sim 0.1$.

In addition to the above considerations, the $\tau \approx 0.45 \; \mathrm{Gyr}$ lifetime for `Oumuamua, as suggested by \cite{2021JGRE..12606706J} to explain its origin near the Local Standard of Rest, implies that the the object originated from a subset of young stars. The current total star formation rate (SFR) of $1.65 \pm 0.19 \; \mathrm{M_\odot \; yr^{-1}}$ implies a source population of $\sim 7 \times 10^8 \; \mathrm{M_\odot}$, namely $(SFR \times \tau) / M_{\star} \sim 1\%$ of the Galactic stellar population where $M_{\star}$ is the Galactic stellar mass \citep{2015ApJ...806...96L, 2019ApJ...878L..11I, 2019A&A...624L...1M, 2021MNRAS.501..302A}, raising the necessary production rate of nitrogen fragments per star by two orders of magnitude. Increasing $\tau$, the age of the nitrogen fragments, would require a larger mass by the same factor, multiplied by an additional factor accounting for the fact that a larger initial surface area implies a greater sublimation rate, which makes the mass budget worse for longer lifetimes.

The mass budget is further increased if erosion of nitrogen icebergs by cosmic rays is properly taken into account. Specifically, \cite{2021arXiv210904494P} showed that the $\tau \approx 0.45 \; \mathrm{Gyr}$ lifetime requires an initial radius of $10 \mathrm{ \; km}$, rather than the $\sim 0.1 \mathrm{ \; km}$ initial radius proposed in  \cite{2021JGRE..12606706J}. We add a cosmic-ray factor $f_{CR}$ that specifies the required increase in the initial mass relative to the $\sim 0.1 \mathrm{ \; km}$ radius. Accounting properly for cosmic-ray erosion, with an initial radius of $10 \mathrm{ \; km}$, would increase the value to $f_{CR} \sim 10^6$, thereby raising the mass budget, $f_{\star}$, described in equation \eqref{eq:limit} by an additional six orders of magnitude. Moreover, \cite{2021arXiv210811194L} showed that less $\mathrm{N_2}$ may actually get ejected from exo-Plutos than assumed by \cite{2021JGRE..12606807D}. If this is the case, then it decreases the efficiency $\epsilon$ and increases $f_{\star}$ further. Conservatively, we ignore this enhancement.

The fraction of heavy elements in stars that must be converted into exo-Plutos in order to explain the detection of `Oumuamua is,
\begin{equation}
f_{\star} = \frac{m \; n \; (M_{\star} / (SFR \times \tau)) \; f_{CR}}{\rho_{\star} (m_{N2} / m_p) f_Z \epsilon \eta} \; \; ,
\end{equation}
or,
\begin{align}
\begin{split}
    \label{eq:limit}
         \log_{10} & {f_{\star}}   \sim   \;  (1.3^{+0.75}_{-1.5})  \log_{10} \left[  \left( \frac{n}{0.1 \mathrm{\; AU^{-3}}}\right) \right. \times \\ & \left( \frac{m}{2.4 \times 10^{11} \mathrm{\; g}} \right) \left( \frac{\rho_{\star}}{0.04 \; \mathrm{M_{\odot} \; pc^{-3}}}\right)^{-1} \times \\ & \left( \frac{M_{\star} / (SFR \times \tau)}{10^2}\right) \left( \frac{D_{N2}}{3 \; \mathrm{km}}\right)^{-1} \times \\ &   \left( \frac{f_Z}{1.3 \%}\right)^{-1} \left( \frac{\epsilon}{0.2}\right)^{-1} \left. \left( \frac{\eta}{1}\right)^{-1} \left( \frac{f_{CR}}{1}\right) \right] \; , 
\end{split}
\end{align}
where the quoted errors are the $95\%$ Poisson error bars for a single detection.


\begin{figure}
 \centering
\includegraphics[width=1\linewidth]{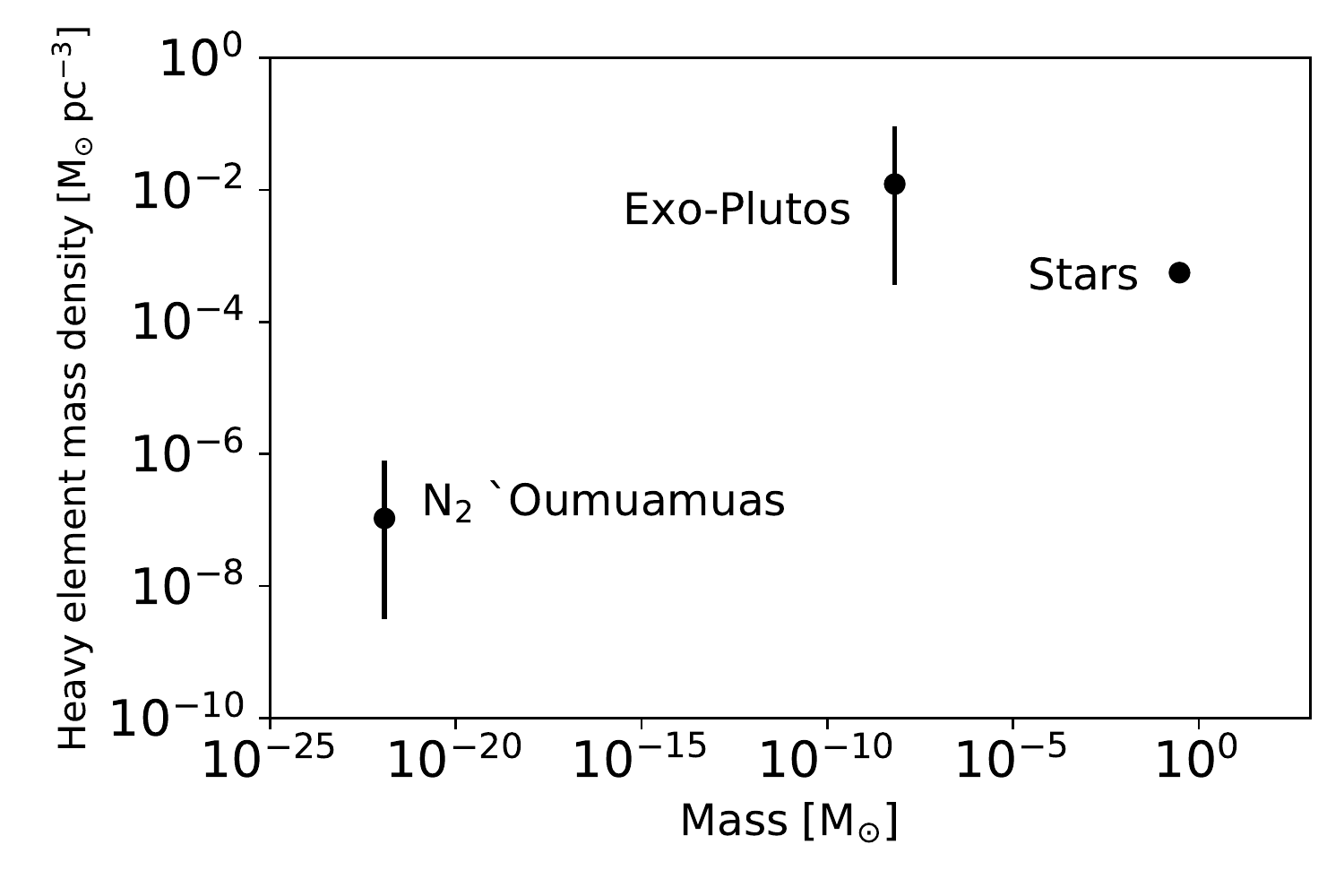}

\caption{Comparison of the mass densities of heavy elements for the $\mathrm{N_2}$ `Oumuamua model (using the mass from \citealt{2021JGRE..12606706J} and the number density from \citealt{2018ApJ...855L..10D} adjusted to reflect the fact that no new `Oumuamua-like objects have been detected in the past few years), the exo-Plutos necessary for producing such $\mathrm{N_2}$ `Oumuamuas (adopting the fiducial values in equation \ref{eq:limit}), and stars \citep{2017MNRAS.470.1360B}. The error bounds on stars are small and not resolvable on this plot.}
\label{fig:mass}
\end{figure}

\begin{figure}
 \centering
\includegraphics[width=1\linewidth]{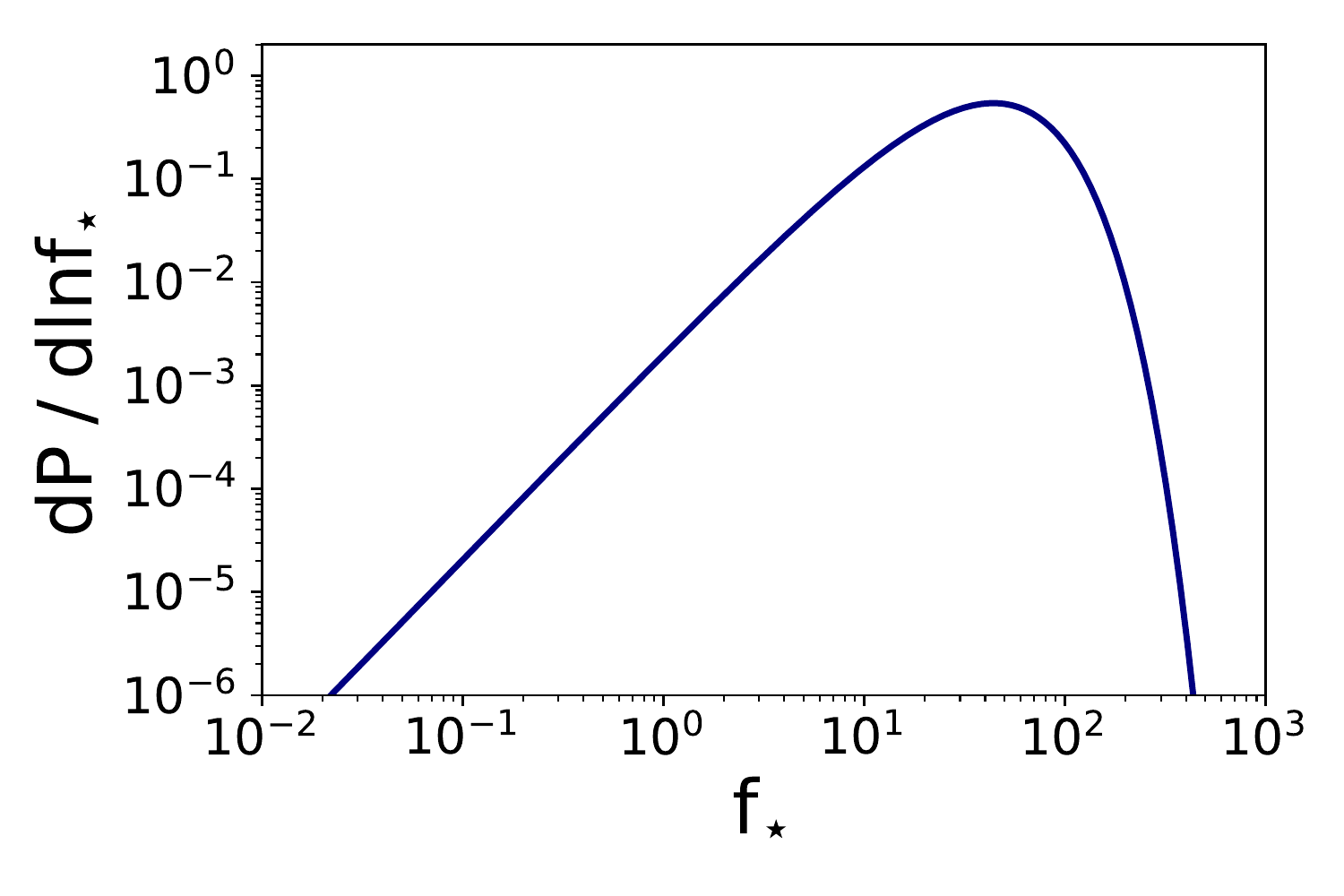}

\caption{The probability distribution for $f_{\star}$, the fraction of heavy elements in stars that must be converted into exo-Plutos to explain the detection of `Oumuamua, adopting Poisson statistics for a single event and the fiducial values in equation \eqref{eq:limit}.}
\label{fig:dpdf}
\end{figure}


\section{Discussion}
\label{sec:d}

Figure \ref{fig:mass} illustrates the mass densities of heavy elements implied for the $\mathrm{N_2}$ `Oumuamua scenario and the exo-Plutos needed to produce such a scenario, with stars shown for reference. The Poisson probability distribution for $f_{\star}$ given the fiducial values in equation \eqref{eq:limit}, namely $dP / d \ln{f_{\star}} = 20^{-2} \; e^{f_{\star}/20} \; f_{\star}^2$, where 20 is the resulting central value for $f_{\star}$ and where $\int_{0}^{\infty} P(f_{\star}) \; df_{\star} = 1$, is displayed in Figure \ref{fig:dpdf}.

The $95\%$ Poisson confidence interval on the fraction of heavy elements in stars required to be converted into exo-Plutos that would produce a population of objects consistent with the detection of `Oumuamua, given the values adopted in equation \eqref{eq:limit}, is $0.7 \lesssim f_{\star} \lesssim 120$. The lower bound of this range would correspond to a mass two orders of magnitude larger than the minimum mass solar nebula \citep{10.1143/PTPS.70.35} being converted exclusively into exo-Plutos. For the number density of fragments to match the central value of $0.1 \mathrm{\; AU^{-3}}$, the mass of material converted into exo-Plutos would typically have to exceed the mass in stars by a factor of $f_{\star} \sim 20$. The nitrogen fragment hypothesis is strongly disfavored, since no known physical process could accomodate such a mass budget. The scenario is untenable because only a small fraction of the mass in stars ends in exo-Plutos in the minimum mass solar nebula model \citep{10.1143/PTPS.70.35}. 

If `Oumuamua were a normal rock, the implied mass budget to produce such a population of objects would still be in tension with our current understanding of planetary systems \citep{2009ApJ...704..733M, 2018ApJ...866..131M, 2019AJ....157...86M}. Note that the mass budget discrepancy discussed here goes far beyond these previous tensions as `Oumuamua is interpreted to be composed of pure nitrogen. The mass budget is reduced considerably if Oumuamua is a thin object, $\lesssim 1 \mathrm{\; mm}$ \citep{2018ApJ...868L...1B}, in which case the total mass of heavy elements needed would be that of a few-kilometer scale asteroid per star \citep{loeb21}.

Additionally, Pluto is primarily composed of rock and water ice, with nitrogen ice comprising just 0.5\%, so the abundance of planetesimals composed of rock and water ice that made the planets should greatly exceed that of nitrogen icebergs. \cite{2021JGRE..12606807D} claim that Pluto may have had a higher nitrogen ice abundance in the past, but do not posit that this abundance exceeded that of rock or of water ice. Even if we generously include a factor of $10^2$ increase in $\mathrm{N_2}$ ice on Pluto-like bodies as suggested by \cite{2021JGRE..12606807D}, the \cite{2021arXiv210904494P} result implies a shortfall in the amount of nitrogen ice by an extra factor of $10^4$. Indeed, long-period comets, originating from the Oort cloud, are known to be composed of mixtures of rock and water ice. Interstellar objects could originate dynamically from stellar perturbations or the Galactic tide on exo-Oort clouds. So, in addition to the unrealistic mass budget, invoking a nitrogen composition for the first interstellar object is also problematic because of its tension with the relative abundance of icy rocks observed in the solar system.

\vspace*{0.3in} 
\section*{Acknowledgements}
We thank Greg Laughlin, Garrett Levine, Ed Turner, and Josh Winn for helpful comments on the manuscript. This work was supported in part by a grant from the Breakthrough Prize Foundation.\newline \newline

\section*{Data Availability}
No new data were generated or analysed in support of this research.




\bibliographystyle{mnras}
\bibliography{example} 





\bsp	
\label{lastpage}
\end{document}